\documentclass[a4paper,10pt,fleqn]{article}
\usepackage{amsmath,amsthm} 
\usepackage[dvipdfmx]{graphicx,color} 
\numberwithin{equation}{section} %equation number--- (2.1)
\author{Yoshio Nishiyama\footnote{e-mail: nisiyama@ynu.ac.jp}\; and Fumiaki Tajima\footnote{e-mail: tajima@ynu.ac.jp} \vspace*{5mm}\\
Yokohama National University, Faculty of Education and Human Sciences,\\
 79-2 Tokiwadai Hodogaya-ku, Yokohama,
240-8501, JAPAN}
\begin{document}

\title{ Trajectory of motion of an electron in the Coulomb scattering in terms of
 the Schr\"{o}dinger wave equation and the Hamilton Jacobi equation}
 
%\date{\today}  
 \maketitle
\begin{abstract}
The trajectory of motion of a scattering electron in the Coulomb potential from the wave function of the Schr\"{o}dinger equation  is presented in two ways, spherical polar coordinates and Temple coordinates, and is compared with each other and with the corresponding motion of classical mechanics.
A good correspondence among dynamics by wave functions and the classical dynamics has been acknowledged by comparing computed examples.
Detailed computing examples discriminate the optimal dynamics of the wave function that should be verified by an experiment.
\end{abstract}
 

\section{Introduction}
We can manipulate an atom to move to where we intend these days.\cite{IBM130501}
Quantum mechanics teaches that the motion of the atom in the region of minute scale should obey the wave equation.

To detect the exact length of e.g. 1 nm it is necessary to measure the fluctuation of the  wave motion reflecting the effect of the 1 nm length.
But the wave length could be far larger than 1 nm.
This has been verified and realized as SNOM [ scanning near field optical microscope]. 
We have shown that the interval of 1 nm can be detected by the visible light of wave length of 441.6nm.\cite{TNOR}
These indicate that the measurement of a matter of length less than the wavelength by the light wave does not obey no diffraction limit nor any indeterminacy.

Molecular dynamics in chemical physics uses trajectories of the concept of classical mechanics to interpret the bond or structure of molecules.\cite{MolDyn}
The concept of trajectory of an atom is useful to understand the structure of aggregates of atoms.

Trials to seek the trajectory in the wave motion had been done, for example,
the trajectory in the Schr\"{o}dinger wave\cite{Bohm} and the ray in the optical diffracted wave~\cite{Keller}.
The concept of trajectory relates closely to the causal interpretation of quantum mechanics.~\cite{Holland}

In what follows we restrict the presentation to the algorithm of the motion 
of an electron in the Coulomb  potential from the wave function
 and do not touch any interpretation about the function or its absolute value.

The hint of derivation of the concept of trajectory from the wave equation is the relation between the electromagnetic wave and the geometrical optics.
The relation between the Maxwell equation and the eikonal equation of geometrical optics has been investigated in detail.~\cite{BWolf1} 
It is well known that the concept of ray, trajectory, derived from the light wave plays practically and theoretically important role.

The eikonal equation in the Schr\"{o}dinger equation is the Hamilton Jacobi equation which is derived by WKBJ approximation to the wave function.
The Hamilton Jacobi equation determines the Hamilton's characteristic function that determines the motion of the particle.\cite{Goldstein}
Thus we should make the mode characteristic function from the wave function that 
can determine the motion of the particle.

In the present paper a trajectory and dynamics of a scattering electron in the Coulomb potential is derived from the wave function described in the spherical polar coordinates
 and another dynamics from the scattering wave function used by Temple and in the text book is also derived.~\cite{Temple,MottMassey} The dynamics for the corresponding motion of the electron in classical mechanics is presented for comparison.

These classical dynamics, dynamics by the wave functions in the spherical polar coordinates and dynamics by the Temple wave functions of a scattering electron are investigated numerically and the difference among them is noted.

In section~\ref{sec:QDWF}  the mode trajectory and 
 dynamics of a particle derived from the wave function in completely separated coordinates system is presented.

 In section~\ref{sec:H-Jeq} dynamics of the scattering electron in the Coulomb
 potential by the Hamilton Jacobi equation in the spherical polar coordinates
 is reviewed briefly. The Hamilton's characteristic function plays the central role 
to derive the orbit and the time elapse of the motion of the electron as is well known.

 In section~\ref{Sec:QMr-x} by following Hamilton's characteristic function of the preceding section we make the mode characteristic function from the wave functions in the spherical polar coordinates and derive the mode trajectory and time elapse of the motion of the electron according to section~\ref{sec:QDWF}.

 In section~\ref{Sec:CMTemple} the Hamilton's characteristic function for the Temple coordinates known in the scattering in quantum mechanics is made to derive the classical motion of the scattering electron by introducing some technical manipulation.
As a result this motion is equivalent to the motion derived in section~\ref{sec:H-Jeq}. 

 In section~\ref{Sec:QMTemple} by using the technique in section~\ref{Sec:CMTemple}  we find out the mode characteristic function from the wave functions in the Temple coordinates and get the mode trajectory and time elapse of the scattering electron. 
The motion of the electron is almost equal to the motion in section~\ref{Sec:CMTemple}. 

 In section~\ref{sec:result} 
dynamics of the scattering electron in the Coulomb potential 
obtained in previous sections \ref{Sec:QMr-x}, \ref{Sec:CMTemple} and \ref{Sec:QMTemple} have been numerically investigated. 
Detail calculation indicates that dynamics in section~\ref{Sec:QMr-x} is reasonable throughout everywhere.
Dynamics in section~\ref{Sec:QMTemple}  shows a defect near the origin of the potential
 while in the other space it is almost equal to the classical dynamics in section~\ref{Sec:CMTemple}.

 In section~\ref{Sec:Conclusion}
 conclusions are described. Dynamics in section~\ref{Sec:QMr-x} should be verified by experiment.

\section{ Wave function and dynamics of an electron}\label{sec:QDWF}
The stationary scattering state wave function consists of travelling waves.\cite{GORDON}
The WKBJ approximation of the travelling wave leads to the Hamilton's characteristic function. We find the mode characteristic function of the travelling wave and define the dynamical equations of the particle in the wave equation.

 The dynamics that leads to the mode trajectory of an electron
 in an attractive Coulomb potential with a charge $Ze ( > 0)$ is
  summarized.~\cite{JOSAA95}
 The wave function  $\Psi$ describing the motion of an electron
  satisfies the Schr\"odinger equation
\begin{align}
 i\hbar \frac{\partial \Psi ({\bf r},t)}{\partial t}
   = \left(  - \frac{\hbar^2}{2m} \triangle  - \frac{Ze^2}{r} \right)
    \Psi ({\bf r},t),
   \label{eq:Schroe}
\end{align}
 where constant $m$ or $-e$ is electron mass or charge, respectively.
 
The equation is assumed to be separable in variables $t, x_1, x_2$
 and $x_3 $.
Let the wave function be
\begin{align}
 \Psi ({\bf r},t) = e^{-iEt/\hbar} \Phi(x_1, x_2, x_3, E, \alpha ,\beta)
 = e^{-iEt/\hbar} \Phi_1(x_1,E,\alpha) \Phi_2(x_2,\alpha ,\beta) \Phi_3(x_3,\beta ),
\end{align}
 where  $E, \alpha$ and $\beta $ are constants of separation, and $E$ is 
 assumed to be the energy of the system.
 These constants should be called mode parameters.
 The wave function of the form 
\begin{align}
   \Phi_j(x_j) =|\Phi_j(x_j)| \exp\{ i \Im \log \Phi_j(x_j) \}
     \equiv |\Phi_j(x_j)| \exp \{ iW_j (x_j) \}, \qquad
     j = 1, 2, 3,
                                    \label{eq:Wjxj}
\end{align}
 is sought, where $\Im$ stands for the imaginary part of, and
 functions $W_j$'s are real.
This should be called a travelling wave where  $W_j$'s satisfy the following.

 Let functions $W_j$'s  satisfy the condition that
  in each classical region of $x_j$ for $j = 1,2,3$
 where classical mechanics hold true for the motion of the particle
 \begin{align}
      W_j(x_j) \simeq  W_j(x_j)_{\rm cl},
  \end{align}
 where the sum of them
 \begin{align}
      W_{\rm cl} = \sum_{j = 1}^3 W_j(x_j)_{\rm cl}
  \end{align}
 is the Hamilton characteristic function of the Hamilton-Jacobi
 equation in classical mechanics.~\cite{Goldstein}
 $W_j(x_j)_{\rm cl}$ is usually obtained as the WKBJ approximation from
 the wave function.
The classical region stands for the domain in which the characteristic
 function holds true.
 
 If  $W_j$'s are found uniquely, the sum of them
 \begin{align}
 W(x_1, x_2, x_3, E, \alpha , \beta ) = W_1(x_1,E, \alpha )
  + W_2(x_2, \alpha ,\beta ) + W_3(x_3, \beta ) 
  = \Im \log \{ \Phi_1(x_1) \Phi_2(x_2) \Phi_3(x_3) \} 
                     \label{eq:Wmcf}
 \end{align}
 is named the mode characteristic function (abbreviated as mcf)
 for the system.~\cite{JOSAA95}
 
By using a general form of the separated functions \eqref{eq:Wjxj}
\begin{align}
 &\Phi_j(x_j;\alpha, \beta, E) 
 = |\Phi_j(x_j;\alpha, \beta, E)| 
    \exp\{ i \Im \log \Phi_j(x_j;\alpha, \beta, E) \}  \notag\\
 &\qquad\equiv 
 |\Phi_j(x_j;\alpha, \beta, E)| \exp \{ iW_j (x_j; \alpha, \beta, E) \},
    \label{eq:Wjxj_ext}
\end{align}
the dynamics of the electron is assumed to be given by
\begin{subequations}
 \begin{align}
 &\frac{ \partial W_j (x_j;\alpha, \beta, E)}{
   \partial \alpha} 
 = \Im \frac{\partial_\alpha \Phi_j(x_j;\alpha, \beta, E)}{
   \Phi_j(x_j;\alpha, \beta, E) } = c_\alpha,   
 \label{eq:trjectry-a} \\
 &  \frac{ \partial W_j (x_j;\alpha, \beta, E)}{
   \partial \beta} 
 = \Im \frac{\partial_\alpha \Phi_j(x_j;\alpha, \beta, E)}{
   \Phi_j(x_j;\alpha, \beta, E) } = c_\beta,   
  	\label{eq:trjectry-b} \\
 &\hbar \frac{ \partial W_j (x_j;\alpha, \beta, E)}{
  \partial E}  = \hbar \Im \frac{\partial_E \Phi_j(x_j;\alpha, \beta, E)}{
   \Phi_j(x_j;\alpha, \beta, E) } = t - t_0.    
      \label{eq:dynamics}
 \end{align}
\end{subequations}
Here $ t_0, c_{\alpha}$ and $c_{\beta}$ are constants (independent of $t$)
 to be determined by initial conditions for the system.
 Equations \hspace{-1mm}\eqref{eq:trjectry-a} and \eqref{eq:trjectry-b} 
 determine the mode trajectory (abbreviated as m-trajectory).
 Variable $t$ of Eq. \hspace{-1mm}\eqref{eq:dynamics} is considered to be
  the dynamical time for the mode trajectory.

\section{ Orbit of an electron in the Coulomb potential
 by Hamilton Jacobi equation in terms of spherical polar coordinates}\label{sec:H-Jeq}
 In the spherical polar coordinates system, $(r, \theta, \phi)$.
the Hamilton characteristic function can be written as follows and satisfies
 the Hamilton Jacobi equation~\cite{Goldstein}
\begin{align}
 &W_{\rm cl}(r, \theta, \phi, E, l ) = W_{r, \rm cl}(r, E, l)
  + W_{\theta, \rm cl}(\theta, l ),    \notag\\
&\frac{1}{2m}\left ( \nabla W_{r, \rm cl}  \right )^2 - \frac{Ze^2}{r} 
 =\frac{1}{2m}\left ( \frac{\partial W_{r, \rm cl}}{\partial r} \right )^2
 + \frac{1}{2m}\frac{1}{r^2}\left ( \frac{\partial W_{\theta, \rm cl}}{
   \partial \theta} \right )^2 - \frac{Ze^2}{r} = E.
   \label{eq:H-Jeq}
\end{align}
$E$ stands for the energy and the charge $Ze$ is attractive for the electron
 if $Z>0$.
We restrict the motion of an electron to the scattering state of $E>0$
 throughout in what follows.
The motion of an electron can be restricted in a plane $(r, \theta)$
 as is well known.
By introducing a variable of separation $L$ standing for the angular
 momentum,  $W_{r, \rm cl}$ and $W_{\theta, \rm cl}$ are determined
 from equations
\begin{align}
  \frac{1}{2m} \left[
   \left(\frac{\partial W_{r, \rm cl}}{\partial r} \right)^2
  + \frac{L^2}{r^2} \right] - \frac{Ze^2}{r} = E,  \\  
   \left(\frac{\partial W_{\theta, \rm cl}}{\partial \theta} \right)^2
    = L^2.
\end{align}
Some calculation gives the results.
\begin{align*}
 &\frac{\partial W_{r, \rm cl}}{\partial r}
   = \sqrt{2mE + \frac{Ze^2}{r} - \frac{L^2}{r^2} }
 = \frac{\sqrt{2mE}}{r} \sqrt{ (r - r_1)(r - r_2) }, \\
 &\qquad    r_{1,2} = - \frac{Ze^2}{2E} \pm \sqrt{\left (\frac{Ze^2}{2E} \right )^2
      + \frac{L^2}{2mE}}. 
 \\
 &\frac{\partial W_{r, \rm cl}}{\partial L}  
  = -2 \arctan \left ( \sqrt{\frac{r/r_1 - 1}{r/( - r_2) + 1}} \right ),  \\
 &\frac{\partial W_{r, \rm cl}}{\partial E} 
  = \sqrt{\frac{m}{2E}} \left (\sqrt{ (r - r_1)(r - r_2)} - \frac{Ze^2}{2E} \log \left |
   \frac{\sqrt{r - r_1}+\sqrt{r - r_2}}{\sqrt{r - r_1} - \sqrt{r - r_2}} \right |
    \right ).
\end{align*}
 The orbit equation  from %with respect to $r, \theta$ is
 $r = \infty, p r \sin \theta = L > 0 (\theta \to \pi) $ to 
 the returning point $ r=r_1, \theta(r_1) $ is
\begin{align}
 & - \frac{\partial W_{r, \rm cl}}{\partial L}
  - \frac{\partial W_{\theta, \rm cl}}{\partial L} 
  = 2 \tan^{-1} \sqrt{\frac{r/r_1 - 1}{r/(-r_2) + 1}} - \theta
  = 2 \tan^{-1} \sqrt{\frac{-r_2 }{r_1}} - \pi
  = \text{const}.     \label{eq:CMobtinc}
\end{align}
 
The returning orbit equation from $r_1, \theta(r_1)$ to $r \to \infty,
 \theta_\text{sc}(r = \infty)$ is
 \begin{align}
 &  \frac{\partial W_{r, \rm cl}}{\partial L}
  - \frac{\partial W_{\theta, \rm cl}}{\partial L} 
  = - 2 \tan^{-1} \sqrt{\frac{r/r_1 - 1}{r/(-r_2) + 1}} - \theta
  = 2 \tan^{-1} \sqrt{\frac{-r_2 }{r_1}} - \pi
    = \text{const. defined at $r=r_1$}.  \label{eq:CMobtsct}
\end{align}
It can be proved that the orbit thus obtained is equivalent to the Temple orbit by classical mechanics \eqref{eq:CMTmpinc}  and \eqref{eq:CMTmpsct}, or \eqref{eq:CMTmpObt}. 

\begin{align}
  \theta_\text{sc} = - 4 \tan^{-1} \left (\sqrt{\frac{-r_2 }{r_1}} \right ) + \pi.
       \label{eq:sctangle_cls}
\end{align}
This expression of the scattering angle is equivalent
 to \eqref{eq:CMTmpsctangl}.

The time elapse of the orbit is
\begin{align}
 &t^\text{in/sc} = \mp \frac{\partial W_{r, \rm cl}}{\partial E} 
 = \mp \sqrt{\frac{m}{2E}} \left ( \sqrt{ (r - r_1)(r - r_2)} 
   - \frac{Ze^2}{2E} \log \left |
   \frac{\sqrt{r - r_1}+\sqrt{r - r_2}}{\sqrt{r - r_1} - \sqrt{r - r_2}} 
  \right | \right )  \notag\\ 
 &\qquad \simeq \mp \sqrt{\frac{m}{2E}} r \text{ as } r \to \infty.
  \label{eq:TimeCL} 
\end{align}
This  is  concordant with Temple time elapse \eqref{eq:CMTmptimeinc} and \eqref{eq:CMTmptimesct}.

\subsection{Cross section}
The differential cross section is expressed in terms of the scattering angle
$ \theta_\text{sc}$ and the impact parameter $s = L/ \sqrt{2mE}$ by
 (3.93) in the textbook \cite{Goldstein}
\begin{align}
  \sigma(\theta_\text{sc}) = \frac{s}{\sin \theta_\text{sc}} \left | 
  \frac{d s}{d \theta_\text{sc}} \right |.   \label{eq:crossecCM}
\end{align}
From \eqref{eq:sctangle_cls} the impact parameter is related to the scattering angle
 as  
\begin{align}
  &s = \frac{Ze^2}{2E} \cot \frac{\theta_\text{sc}}{2}.
\end{align}
The differential cross section is 
\begin{align}
  & \sigma(\theta_\text{sc}) = \frac{1}{4}  \left ( \frac{Ze^2}{2E} \right )^2
     \csc^4 \left (\frac{\theta_\text{sc}}{2} \right )
   = \frac{ \eta_s^2}{4 k^2} \csc^4 \left (\frac{\theta_\text{sc}}{2} \right ).
          \label{eq:crossecCM1} 
\end{align}
This is the same as \eqref{eq:crossect_templ}  where $k$ and $\eta_s$ is determined in \eqref{eq:etasrho}.

\section{ Mode trajectory of an electron by the wave function
 in terms of spherical polar coordinates  \label{Sec:QMr-x}}
 The scattering state of an electron in the Coulomb potential is analyzed
 in the spherical polar coordinate system.
 The wave function $\Psi ({\bf r},t)$ is expressed in the spherical
 polar coordinates with mode parameters,
 constants of separation of variables, 
  $E, \nu$  and $\mu $ as
\begin{align}
  \Psi({\bf r},t) &= \exp \left( -i E t / \hbar \right)
   \Phi ({\bf r},E), \label{eq:Psirt} \\
  \Phi ({\bf r},E) &= R(r,E,\nu )Y(\theta ,\nu , \mu )
                     \exp(i\mu \phi ).  \label{eq:PhirEnm}
\end{align}
Constant $E$ stands for the energy and $\hbar \nu$  for the orbital
 angular momentum, and $\hbar \mu$  represents the component of
 the angular momentum along the polar axis.
When $\nu$ and $\mu$ are integral numbers, they are usual azimuthal and
 magnetic quantum number.~\cite{SchiffS4}

In what follows $\mu=0$ is assumed.
  $Y(\theta ,\nu , 0 )$ is written as $Y_\nu(\theta )$.

The mcf expressed in terms of the spherical polar coordinates are obtained 
 as follows.
The function $Y_\nu(\theta )$ satisfies the differential equation
\begin{align}
\left[ \frac{d^2}{d\theta ^2} + \cot \theta \frac{d}{d\theta }
 + \nu (\nu +1)  \right]  Y_\nu(\theta ) = 0.
                                   \label{eq:Ytheta}
\end{align}
The solution is a linear combination of linearly independent
 associated Legendre functions,
 $ P_{\nu} (\cos \theta )$ and
  $ Q_{\nu} (\cos \theta )$.~\cite{Erdely} By putting $x=\cos \theta$ \\
\begin{align}
 & P_{\nu}(x) = F \left( -\nu, \nu+1; 1; \frac{1-x}{2} \right ),
   \label{eq:Pnu_x}  \\
 &  Q_{\nu}(x)
  =  \pi^{\tfrac{1}{2}} 
  \Biggl \{ - \frac{\Gamma(1/2+\nu/2 )}{2\Gamma(1+\nu/2 )}
  \sin \{ \tfrac{\pi}{2} \nu \}
  F \left (-\frac{\nu}{2}, \frac{1+\nu }{2}; \frac{1}{2}; x^2 \right ) 
  \notag\\
  &\qquad \hspace{3cm}
  +  \frac{x \Gamma(1+\nu/2 )}{\Gamma(1/2+\nu/2 )}
  \cos ( \tfrac{\pi}{2} \nu )
  F \left (\frac{1-\nu }{2}, 1+\frac{\nu }{2}; \frac{3}{2}; x^2 \right )
  \Biggl \}.    \label{eq:Qnu_x}
\end{align}
\begin{align}
 &\frac{\partial }{\partial \nu} P_{\nu}(x)
 =  \left \{ - \frac{\partial }{\partial a} 
              + \frac{\partial }{\partial b} \right \}
   F\left( a, b; 1; \frac{1-x}{2} \right )_{a=-\nu, b=\nu+1 },
  \label{eq:Pndnu}   \\
 &\frac{\partial }{\partial \nu} Q_{\nu}(x)
  = \pi^{\tfrac{1}{2}}  \frac{1}{2} 
  \Biggl [ - \frac{\Gamma(1/2+\nu/2 )}{2\Gamma(1+\nu/2 )}
  \sin ( \tfrac{\pi}{2} \nu ) \Biggl \{ \psi(1/2+\nu/2 )
  - \psi(1+\tfrac{1}{2}\nu )    \notag \\ 
 &\quad \hspace{4cm}+ \pi \cot (\tfrac{\pi}{2} \nu )
  - \frac{\partial}{\partial a} + \frac{\partial}{\partial b} \Biggl \}
  F \left (a,b; \frac{1}{2}; x^2
        \right )_{a=-\tfrac{1}{2} \nu, b=\tfrac{1}{2}(1+\nu )}
    \notag \\
  &\qquad  \hspace{1cm}
  +  \frac{x \Gamma(1+\nu/2 )}{\Gamma(1/2+\nu/2 )}
  \cos ( \tfrac{\pi}{2} \nu )\Biggl \{ \psi(1 +\nu/2 )
  - \psi(1/2+\nu/2 )   \notag\\
 &\quad \hspace{4cm} - \pi \tan (\tfrac{\pi}{2} \nu )
  - \frac{\partial}{\partial a} + \frac{\partial}{\partial b}
   \Biggl \} 
  F \left (a, b; \frac{3}{2}; x^2
   \right )_{a=\tfrac{1}{2}(1-\nu ), b=1+\tfrac{1}{2} \nu }
  \Biggl ].       \label{eq:Qndnu}
\end{align}
$F(a,b;c;z)$ is the hypergeometric function usually written as
 ${}_2F_1(a,b;c;z)$.
\begin{align}
 &\frac{\partial^2 }{\partial \nu^2} P_{\nu}(x)
 =  \left ( - \frac{\partial }{\partial a} 
              + \frac{\partial }{\partial b} \right )^2
   F\left( a, b; 1; \frac{1-x}{2} \right )_{a=-\nu, b=\nu+1 }, 
      \label{eq:Pnddnu}   \\
 &\frac{\partial^2 }{\partial \nu^2} Q_{\nu}(x)
  =  \frac{\sqrt{\pi}}{4} 
  \Biggl [ - \frac{\Gamma(1/2+\nu/2 )}{2\Gamma(1+\nu/2 )}
  \sin ( \tfrac{\pi}{2} \nu ) \Biggl \{ \psi'(\tfrac{1}{2}+\tfrac{1}{2}\nu )
  - \psi'(1+\tfrac{1}{2}\nu ) - \pi^2 \csc^2( \tfrac{\pi}{2} \nu ) \notag \\ 
 &\quad + \left ( \psi(\tfrac{1}{2}+\tfrac{1}{2}\nu )
  - \psi(1+\tfrac{1}{2}\nu ) + \pi \cot (\tfrac{\pi}{2}\nu) \right )^2
 +2 \left ( \psi(\tfrac{1}{2}+\tfrac{1}{2}\nu  )
  - \psi(1+\tfrac{1}{2}\nu ) + \pi \cot (\tfrac{\pi}{2}\nu) \right )
       ( - \partial_a + \partial_b )
   \notag \\ 
 &\quad + ( - \partial_a + \partial_b )^2 
  \Biggl \}  F \left (a,b; \frac{1}{2}; x^2
        \right )_{a=-\tfrac{1}{2} \nu, b=\tfrac{1}{2}(1+\nu )}
    \notag \\
  &+ \frac{x \Gamma(1+\nu/2 )}{\Gamma(1/2+\nu/2 )}
  \cos ( \tfrac{\pi}{2} \nu )\Biggl \{ \psi'(1 +\tfrac{1}{2}\nu )
  - \psi'(\tfrac{1}{2}+\tfrac{1}{2}\nu ) - \pi^2 \sec^2 (\tfrac{\pi}{2} \nu )
  \notag\\
 &\quad +\left (\psi(1 +\tfrac{1}{2}\nu ) - \psi(\tfrac{1}{2}+\tfrac{1}{2}\nu )
  - \pi \tan (\tfrac{\pi}{2} \nu ) \right )^2
  + 2\left (\psi(1 +\tfrac{1}{2}\nu ) - \psi(\tfrac{1}{2}+\tfrac{1}{2}\nu )
  - \pi \tan (\tfrac{\pi}{2} \nu ) \right )(- \partial_a + \partial_b)
  \notag\\
 &\quad +(- \partial_a + \partial_b)^2 \Biggl \} 
  F \left (a, b; \frac{3}{2}; x^2
   \right )_{a=\tfrac{1}{2}(1-\nu ), b=1+\tfrac{1}{2} \nu }
  \Biggl ].       \label{eq:Qnddnu}
\end{align}
 Equations~\eqref{eq:Pnddnu}  and \eqref{eq:Qnddnu} will be used in \eqref{eq:dnudtetsc}.

A travelling wave in the $\theta$ coordinate space is given as
 by using \eqref{eq:Pnu_x} and \eqref{eq:Qnu_x}
\begin{align}
 Y_\nu(\cos \theta) &\equiv Q_{\nu} (\cos \theta ) + i\frac{\pi}{2}P_{\nu} (\cos \theta )
    \notag\\
    &= | Y_\nu(\cos \theta)| \exp(i \arg  Y_\nu(\cos \theta))
   \equiv  | Y_\nu(\cos \theta)| \exp(i W_{\theta}).
                                     \label{eq:Yteta}
\end{align}

The mcf for the $\theta$ component should be determined as
\begin{align}
 W_{\theta} (\theta ,\nu ) = \Im \log  Y_\nu(\cos \theta)
  = \arctan \left[ \frac{\pi}{2}
  \frac{P_{\nu} (\cos \theta )}{Q_{\nu} (\cos \theta )}
   \right],                 \label{eq:Wntheta}
\end{align}
because of the similarity to the characteristic function 
$W_{\theta, \rm cl}$ in the classical region and the validity of 
the results derived from this as will be seen in the following.
\begin{align}
 \frac{\partial }{\partial \nu} W_{\theta} (\theta ,\nu )
  = \Im \frac{\partial_\nu  Y_\nu(x)}{ Y_\nu(x)}
  = \frac{\pi}{2} \frac{\partial_\nu P_{\nu}(\cos \theta ) Q_{\nu}(\cos \theta )
   - P_{\nu}(\cos \theta ) \partial_\nu Q_{\nu}(\cos \theta ) }{
   Q_{\nu}^2(\cos \theta ) + \tfrac{1}{4}P_{\nu}^2(\cos \theta ) }.
        \label{eq:Wdnutheta}
\end{align}
By the asymptotic expansion of the Legendre functions for
 $\nu \gg 1$,~\cite{AbSt8} it can be obtained that 
 \begin{align}
  W_{\theta}(\theta, \nu) \approx
    \left( \nu + \tfrac{1}{2} \right) \theta + \tfrac{1}{4} \pi,
 \mbox{\hspace*{.5cm} ($\epsilon < \theta < \pi - \epsilon,\; 
   \epsilon > 0$)}.
			\label{eq:Wthetaprx}
\end{align}
We can recognize by numerical calculation that
 $ \partial_\nu W_{\theta}(\theta, \nu) \simeq \theta $ holds true 
  for $\nu+ \tfrac{1}{2} >0$. 

 The value of $W_{\theta}$ at the singular points $\theta = 0$ or $ \pi$ 
 are defined by the ratio of the limiting behaviour of the both Legendre
 functions as follows. ~\cite{Erdely} 
\begin{align}
   W_{\theta}(0, \nu ) =0, \quad
  W_{\theta}(\pi, \nu ) = \pi \nu.   \label{eq:Wtheta0pi}
\end{align}

Behaviour of the Legendre functions near the singular points shows at $x=-1$ ~\cite{Erdely}
\begin{align}
 &Y_\nu(x) = Q_\nu(x) + i \frac{\pi}{2} P_\nu(x)
  \simeq  \frac{ e^{i \nu \pi} }{2} \left [ \Phi(x,\nu) + i\pi \right ],  \notag\\
 & \Phi(x,\nu) \equiv \log(1/2+x/2) + \gamma + 2\psi(\nu+1)
 \simeq 2 \log (\pi/2-\theta/2),
    \label{eq:FAIxn} \\
 &\partial_\nu Y_\nu(x) \simeq i\pi Y_\nu(x) + e^{i \pi \nu} \psi'(\nu+1),
      \notag\\
 & \partial_\nu W_\nu(x) = \Im \frac{\partial_\nu Y_\nu(x)}{Y_\nu(x)}
 \simeq \pi + \Im \left ( \frac{2 \psi'(\nu+1)}{\Phi(x,\nu) + i \pi }
 \right )  \to \pi.\;
  \label{eq:dnuWtetpi}
\end{align}
 
\begin{figure}[htbp]
\begin{center}
 \includegraphics[width=6.5cm,clip]{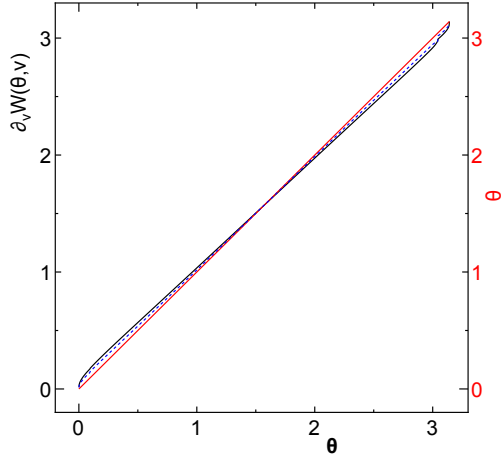} 
 \caption{\small $\partial_\nu W_\theta(\theta, \nu)$ vs $\theta, \nu  = 0.5$(solid line),
 $\nu  = 1.2$(dot line) and $\theta$ vs $\theta$(red line).
\label{fig:1} }
\end{center}
\end{figure}

The graphical example of $ \partial_\nu W_{\theta}(\theta, \nu)$ vs. 
 $\theta $ for $\nu=0.5$ and $\nu=1.2$ with a graph of $\theta$ vs $\theta$
 is shown in Fig.~\ref{fig:1}.

 Radial wave function satisfies the differential equation; 
cf. Classical eq. (10.75) in Goldstein~\cite{Goldstein} 
 $\alpha_\theta^2(=\ell^2) \Leftrightarrow \nu(\nu+1)$
\begin{align}
 \left[ \frac{d^2}{dr^2} - \frac{\nu (\nu +1)}{r^2} + \frac{2m}{\hbar ^2}
 \left( \frac{Ze^2}{r} + E \right) \right] u(r) = 0,
                               \label{eq:urEnu}    
\end{align}
where $u(r,E,\nu ) = r R(r,E,\nu )$. By putting
\begin{align}
 &\eta_s= \frac{Zme^2}{\hbar^2 k}
  = \frac{Ze^2}{\hbar c}\sqrt{\frac{mc^2}{2E}}, \;
 k =  \frac{\sqrt{2mE}}{\hbar},\; kr =\rho,   \label{eq:etasrho} \\
 &\left ( \frac{d^2}{d\rho^2} + 1 + \frac{2\eta_s}{\rho}
   - \frac{\nu(\nu+1)}{\rho^2} \right ) u(\rho) = 0.
	\label{eq:u_rhoeta}  
 \end{align}
 With $E$ positive the linearly independent solutions are
\begin{align}
 u_M &= e^{-i\rho} \rho^{\nu +1} M(\nu +1+i\eta_s ,2\nu +2, i2\rho),
                                     \label{eq:uM} \\
 u_V &= e^{-i\rho} \rho ^{\nu +1} V(\nu +1+i\eta_s ,2\nu +2, i2\rho),
                                     \label{eq:uV}
\end{align}
where $u_M = u_M^*$ is real~\cite{AbraSteg1314}.

 Function $V(a,b,z)$ is defined for convenience~\cite{JOSAA95}
\begin{align}
 V(a,b,z)
  &= \Gamma(a) \left[ U(a, b, z)
   - \cos \pi a \frac{\Gamma(b-a)}{\Gamma(b)} M(a, b, z) \right]
	   					\nonumber\\
  &= -\cos \pi b \frac{\Gamma(1-b)\Gamma(b-a)}{\Gamma(1-a)} M(a,b,z)  
       + \Gamma(b-1) z^{1-b} M(1+a-b,2-b,z).   \label{eq:defV}
\end{align}
 Functions $M(a,b,z)$  and $U(a,b,z)$ are the Kummer
  functions.~\cite{AbraSteg1314}
 
 For the far region from the center of the potential, $\rho  \gg 1$, 
$b, a $ fixed where $a = \nu  + 1 + i\eta_s $ and $b = 2\nu  + 2,$
  it holds ~\cite{AbraSteg1314,Slater} that
\begin{align}
 M(a, b, i2\rho) &\simeq  e^{i\rho} \Gamma (b) \left[
 \frac{e^{-i(\rho - \pi a/2)}}{\Gamma (a^*)}(2\rho )^{-a}
   \left(1 + \frac{i a(1 - a^*)}{2 \rho} \right)
                                             + c.c. \right], \\
V(a,b,i2\rho ) &\simeq e^{i\rho} \left[ -i \sin (\pi a) G(\rho ,a)
 - \cos (\pi a) G(\rho ,a)^{\ast} \right],
\end{align}
where
\begin{align}
 G(\rho ,a) = \Gamma (a)e^{-i(\rho - \pi a/2)}(2\rho )^{-a}
      \left(1 + \frac{i a(1 - a^*)}{2 \rho} \right),
\end{align}
and $G(\rho ,a)^*$ is the complex conjugate (c.c.) of $G(\rho ,a)$.
These asymptotic forms indicate that  the linear
combination of functions $M$ and $V$ producing an outgoing travelling wave 
in the far region from the origin should be written as 
\begin{align}
&u(r, E, \nu) = \exp(-i\rho) \rho^{\nu +1} \left[ V(a, b, i2\rho)
  + i M(a, b, i2\rho) \sin (\pi a)
  \frac{\Gamma (a)\Gamma (b-a)}{\Gamma (b)} \right]  \notag\\
 &\quad= \exp(-i\rho) \rho^{\nu +1} \Biggl [ - \exp(- i\pi b)
  \frac{\Gamma (1-b)\Gamma (b-a)}{ \Gamma (1-a)}  M(a, b, i2\rho) \notag\\
 &\quad+ \Gamma(b-1) (i2\rho)^{1-b} M(1+a-b, 2-b, i2\rho) \Biggr ] 
  \equiv \exp(-i\rho) \rho^{\nu +1} \tilde{u}(\rho, E, \nu).
			\label{eq:u_wave}
\end{align}
By equations mentioned above, this leads to the diverging spherical wave
\begin{align}
 u &\approx   e^{\tfrac{1}{2}\pi \eta_s} 2^{-1-\nu}
   \Gamma(a^*) \exp[i \{ \rho + \eta_s \log (2\rho)  
    - \tfrac{1}{2}\pi (1 + 3 \nu) \} ] 		\nonumber\\
 &\quad \times \left( 
   1 + i \frac{\nu(1 + \nu) + \eta_s^2}{2\rho} - \frac{\eta_s}{2\rho}
     + O(\rho^{-2}) \right),		\label{eq:uLarger}
\end{align}
for $\rho $ large~\cite{GORDON, MottMassey}.

\begin{align}
 &\partial_{\nu} u= \log( \rho) u +  e^{-i\rho} \rho^{\nu +1} \Biggl [
 - e^{- i \pi b} \frac{\Gamma (1-b)\Gamma (b-a)}{ \Gamma (1-a)} \Biggl \{
   - i 2\pi + \psi(1-a) - 2\psi(1-b) + \psi(b-a)    \notag\\
 &\qquad{} + \frac{\partial}{\partial a} 
   + 2\frac{\partial}{\partial b} \Biggr \} M(a, b, i2\rho)  \notag\\
 &\quad{} +  \Gamma(b-1) (i2\rho)^{1-b} \Biggl \{ 2\psi(b-1) 
  -2 \log(i2\rho) - \frac{\partial}{\partial A} - 2\frac{\partial}{\partial B}
  \Biggr \} M(A, B, i2\rho)_{A=1+a-b,B=2-b}   \Biggr ],  
  \label{eq:dnuU}  \\
 &\partial_{\nu} u \approx \Biggl \{ -i \frac{3\pi}{2} - \log(2) +\psi(a^*) 
  + i \frac{\nu + \tfrac{1}{2}}{\rho}\left ( 
   1 + i \frac{\nu(1 + \nu) + \eta_s^2}{2\rho} - \frac{\eta_s}{2\rho} 
   \right )^{-1} \Biggr \} u.
\end{align}

 Eq.~\eqref{eq:uLarger} would suggest that  the travelling 
wave in the $r$ coordinate space should be given by
\begin{align}
 u &= u_V + iu_M \sin (\pi a) \frac{\Gamma (a)\Gamma (b-a)}{\Gamma (b)}
   = |u| e^{i \Im \log(u)}
 \equiv e^{-i\pi \nu } (u_{\text R} + iu_{\text I})  
   \equiv |u| e^{i (W_r - \pi \nu)},
				   \label{eq:wfmotion}
\end{align}
and thus the mcf in the $r$ coordinate is given by
\begin{align}
 W_r(r,E,\nu ) = \arctan \frac{u_{\text I}}{u_{\text R}} = \arg(u) + \pi \nu
   = \Im \log(u) + \pi \nu.
                     \label{eq:Wr}
\end{align}
Here, functions $u_{\text R}$ and $u_{\text I}$ are proved  to be real.

In the far region from the origin the mcf is approximated  as
\begin{align}
  & W_r(r, E, \nu ) \approx \rho  + \eta_s \log 2\rho
      - \arg \Gamma(a) - {}^1\!/\!_2 \pi(1 + \nu)
       + \frac{\nu(1 + \nu) + \eta_s^2}{2 \rho}  + O( \rho^{-2}),
   \label{eq:Wrfar}  \\
 & \partial_\nu W_r(r, E, \nu) \approx - \Im \{ \psi(a) \} - \frac{\pi}{2}
     + \frac{2\nu+1 }{2 \rho}.  \label{eq:dnuWrfar}  
\end{align}
 This is nearly equal to the corresponding Hamilton characteristic 
  function.~\cite{GORDON}
  It indicates the validity of the definition of the mcf \eqref{eq:Wr}.

For $\rho$ small, it is obtained from \eqref{eq:u_wave}, \eqref{eq:dnuU},
 \eqref{eq:Wr} that by using $z=i2\rho$ and
 \begin{align}
 & M(a,b,z) \approx 1,\; \partial_a M(a,b,z) \approx 0,\,
  \partial_b M(a,b,z) \approx 0, \notag\\
 &\tilde{u} \approx 
  - e^{- i \pi b} \frac{\Gamma (1-b)\Gamma (b-a)}{ \Gamma (1-a)}
  + \Gamma(b-1) z^{1-b}, \notag\\
 & W_r(0, E, \nu) = \Im \log \Gamma( b-1) - \tfrac{1}{2} \pi= - \tfrac{1}{2} \pi, \\
 & \partial_\nu W_r(0, E, \nu)  =0.
 \end{align}

For example, a trajectory of an electron incident from a starting point
 distant from the origin of the potential,
 $(\rho_{\text{st}}, \theta_{\text{st}}, \phi_\text{st})$, and
 scattered to another distant scattered point
 $(\rho_\text{sc}, \theta_\text{sc}, \phi_\text{sc})$ is considered.
To be specific, that  $\rho_\text{st} = \infty$ and $\theta_\text{st} = \pi$
 is  assumed for $\nu+1/2 > 0$.
For the m-trajectory from $\rho_\text{st}$ to the origin or
 the returning point $\rho_\text{ret}$, the mcf for descending $\rho$
 and $\theta$ is written as, like the classical H-Jacobi characteristic function 
\begin{align}
  W(r, \theta, E, \nu ) =  - W_r(r, E, \nu ) - W_{\theta}(\theta, \nu ).
		  \label{eq:mcfsctin}
\end{align}
The trajectory is given by the equations \eqref{eq:trjectry-a}, \eqref{eq:trjectry-b}, 
 \eqref{eq:Wntheta} and \eqref{eq:Wr} and by assuming
 $\partial_\nu W_{\theta}(\theta_\text{st}, \nu )=\pi$ for 
 $\nu+1/2>0$,
\begin{align}
 \frac{\partial}{\partial \nu} \left \{- W_r(r, E, \nu )
           - W_{\theta}(\theta, \nu ) \right \} 
 &=  -\partial_\nu W_r(\infty, E, \nu )
      - \partial_\nu W_{\theta}(\theta_\text{st}, \nu ) 
   = \Im \psi(a) -\tfrac{1}{2}\pi 
     \label{eq:trjecQinc} \\
 &=  - \partial_\nu W_r(\rho_\text{ret}, E, \nu )
      - \partial_\nu W_{\theta}(\theta_\text{ret}, \nu ).  
\end{align}
Let $\rho_\text{ret}=0$ then $\partial_\nu  W_r(0) =0 $ for $\nu+1/2>0$, thus
\begin{align}
  \partial_\nu  W_\theta(\theta_\text{ret})
   =\partial_\nu W_r(\infty) + \partial_\nu W_{\theta}(\theta_\text{st}, \nu )
   - \partial_\nu  W_r(0)
    =- \Im \psi(a) + \tfrac{1}{2}\pi.  \label{eq:tet_ret}
\end{align}

For the path from the origin or  the returning point ($\rho_\text{ret}, \theta_\text{ret}$) 
 to the scattered point $(\rho_\text{sc} = \infty, \theta_\text{sc})$,
the mcf for increasing $\rho$ and descending $\theta$ is given by
\begin{align}
  W(r, \theta,   E, \nu ) =  W_r - W_{\theta}.	\label{eq:r0torpi}
\end{align}
The trajectory should be taken to be continuous to the incident trajectory
   at  the returning point ($0, \theta_\text{ret}$). \\
The trajectory equation is written as
\begin{align}
 \frac{\partial}{\partial \nu} \left( W_r - W_{\theta}  \right)
 &=  \partial_\nu W_r(\rho_\text{ret}) 
    - \partial_\nu W_{\theta}(\theta_\text{ret})  
  = \Im \psi(a) - \tfrac{1}{2} \pi   
    \label{eq:trjecQsct} \\
 &=  \partial_\nu W_r(\infty) 
    - \partial_\nu W_{\theta}(\theta_\text{sc}). 
\end{align}

Since function $\partial_{\nu} W_r$ shows monotonic decrease
with respect to $\rho$ while $\partial_{\nu} W_{\theta}$ does
 monotonic increase with respect to $\theta$
 as proved by numerical calculations,
there is a point $\rho = \rho_{\pi}$ where $\theta$ takes $\pi$.
  It can be $\rho_\text{ret} = 0$.
$\partial_{\nu} W_r(\rho=0) = 0$ for $\nu+1/2>0$.
The scattering angle $\theta_\text{sc}$ is given by
\begin{align}
 \partial_{\nu} W_\theta(\theta_\text{sc}, \nu) = - 2 \Im \psi(a).
  \label{eq:scAngle_nu}
\end{align}

The scattering angle $\theta_\text{sc}$ of an incident beam as a function
 of the impact parameter $ks$ for the classical orbit and the parameter $\nu+1/2$ for the mode-trajectory is shown in Fig.~\ref{fig:2} in \S~\ref{sec:result}.
 
\subsection{\sc dynamics time-dependence}
Eq.~\eqref{eq:dynamics} leads to the dynamics along the trajectory.
 Since $\partial W_\theta / \partial E = 0, \partial W / \partial E =
  \partial W_r / \partial E $  of \eqref{eq:Wr}.
From \eqref{eq:etasrho}, \eqref{eq:u_wave}, \eqref{eq:Wr}
\begin{align}
 &\frac{\partial \rho}{\partial E} = \frac{\rho}{2E},\;
   \frac{\partial \eta_s}{\partial E} = - \frac{\eta_s}{2E},\;
 \frac{\partial a}{\partial E} = - \frac{i\, \eta_s}{2E}, \notag\\
 &\frac{t + t_0}{\hbar} 
  = \frac{\partial W_r}{\partial E }
  = \frac{1}{1 + (u_{\text I}/u_{\text R})^2}
   \frac{\partial}{\partial E } \left(
   \frac{u_{\text I}}{u_{\text R}} \right)
  =  \Im \left ( \frac{\partial_E  u}{u } \right ). 
\end{align}

For $\rho \to \infty$ from \eqref{eq:Wrfar} 
\begin{align}
&\frac{\partial }{\partial E} W_r(r, E, \nu) 
  \simeq \frac{\partial \rho}{\partial E} \left (
  1 + \frac{\eta_s}{\rho} \right )
   + \frac{\partial \eta_s}{\partial E} \log (2\rho)
 - \Im \left (\frac{\partial a}{\partial E} \psi(a) \right )
  + \frac{\partial \eta_s}{\partial E} \frac{\eta_s}{\rho},   \notag\\
 & 2E \frac{\partial }{\partial E} W_r(r, E, \nu) 
  \simeq \rho  - \eta_s \log (2\rho)+\eta_s + \eta_s\Re \psi(a) 
  - \frac{\eta_s^2}{\rho}. 
\end{align}
\begin{align}
 &2E \frac{\partial u}{\partial E} = \left ( - i \rho+ \nu + 1 \right ) u
   + e^{-i \rho} \rho^{\nu + 1} \Biggl [ - e^{- i\pi b}
  \frac{\Gamma (1-b)\Gamma (b-a)}{ \Gamma (1-a)} \Biggl \{
  i \eta_s \left \{ \psi(b-a) - \psi(1-a) \right \}
 \notag\\
 &\quad{} -i \eta_s \frac{\partial}{\partial a}
  + z  \frac{\partial}{\partial z} 
  \Biggr \} M(a, b, z)_{z=i2\rho}                  \notag\\
  &\quad{} + \Gamma(b-1) (i2\rho)^{1-b} \Biggl \{ 
    1-b - i \eta_s \frac{\partial}{\partial A}
  + z  \frac{\partial}{\partial z} 
  \Biggr \}  M(A,2-b,z)_{A=1+a-b, z=i2\rho} \Biggr ], 
      \label{eq:2EdudE}  \\
& \frac{2E}{\hbar}(t + t_0) =2E \Im \left (\frac{\partial_E u}{u} \right )
  =2E  \left (\frac{\partial W_r(r,E,\nu)}{\partial E} \right ).
   \label{eq:timelapse}
\end{align}

Eq.~\eqref{eq:2EdudE}  leads to as $\rho \to 0$, 
\begin{align}
 &2E \frac{\partial \tilde{u}}{\partial E}(\rho )
   \approx  
  - \Gamma(b) (i2\rho)^{1-b},   \notag \\
&2E  \left (\frac{\partial W_r(0,E,\nu)}{\partial E} \right ) 
  =2E \Im \left (\frac{\partial_E \tilde{u}(\rho=0)}{\tilde{u}(\rho=0)}
         \right )
 = - \Im(b-1) = 0.
\end{align}

 Incident mcf \eqref{eq:trjecQinc} and returning mcf \eqref{eq:r0torpi}
 leads to the time elapse equation
\begin{align}
 & \frac{2E}{\hbar}(t(r) + t_0)^\text{in} 
 =2E  \left ( - \frac{\partial W_r(r, E, \nu)}{\partial E} \right ),\notag\\
 & \frac{2E}{\hbar}(t(r=0) + t_0)^\text{in} 
  =2E  \left ( - \frac{\partial W_r(0, E, \nu)}{\partial E} \right )
  = 0,          \notag\\
 & \frac{2E}{\hbar}(t(r) + t_0)^\text{sc} 
 =2E \left ( \frac{\partial W_r(r, E, \nu)}{\partial E} \right ).
  \label{eq:timeelapse}
\end{align}

Examples of the mode trajectory and time elapse of a scattering electron are drawn in
 Figs.~\ref{fig:4} and~\ref{fig:5} in \S~\ref{sec:result}.  

\subsection{Cross section}
In the remote region from the origin, equation (\ref{eq:dnuWrfar}) shows that
 the difference between two positions along a trajectory satisfies
\begin{eqnarray}
  \frac{d \theta}{d \rho} = \frac{\partial}{\partial \rho}
  \frac{\partial}{\partial \nu} W_r = - \frac{\nu + \tfrac{1}{2}}{\rho^2}.
\end{eqnarray}
Integration gives rise to $\rho (\pi - \theta) = \nu + \tfrac{1}{2}$ for
 $\rho \to \infty,\; \theta  \to  \pi$.
It is thus obtained that the impact parameter of the trajectory
 is given by
\begin{eqnarray}
  s = r \sin(\pi - \theta)  = \frac{\rho}{k} (\pi - \theta)
    =\frac{\nu + \tfrac{1}{2}}{k}.
\end{eqnarray}
This indicates that $(\nu + \tfrac{1}{2}) \hbar$ corresponds to
 $ l \hbar$, angular momentum in the sense of classical mechanics,
 and $\nu$ should be greater than $- \tfrac{1}{2}$.

More strictly speaking for the m-trajectory in the remote region
 $\rho \to \infty, \theta \to \pi, \text{ or } x \equiv \cos \theta \to -1$,
by using  \eqref{eq:dnuWtetpi} and \eqref{eq:dnuWrfar},
 equation~\eqref{eq:trjecQinc} gives rise to
\begin{align}
 &  \partial_\nu W_r(r, E, \nu ) \simeq  - \Im \psi(a) - {}^1\!/\!_2 \pi(1)
       + \frac{\nu + 1/2}{ \rho},\;
  \partial_\nu W_\nu(x)
 \simeq \pi - \frac{2\pi \psi'(\nu+1)}{\Phi(x,\nu)^2 + \pi^2 },  \notag\\
 & \nu + \frac{1}{2} \simeq \rho  \frac{\pi \psi^\prime(\nu+1)}{
  2 \log^2 (\pi/2 - \theta/2) }.
\end{align}
Therefore $\nu + \frac{1}{2}$ does not exactly stand for
 $\rho \sin (\pi - \theta)$ or the (classical) impact parameter.
By numerical calculation, however, Figure~\ref{fig:2} indicates that $\nu + \frac{1}{2}$  corresponds well to the impact parameter $ks$.

That the height at the starting point $\rho \sin(\pi -\theta) \to 0$ as $\rho \to \infty$ means that m-trajectories seem to start from points of height 0 but they are discriminated by the difference of $\nu$.

The differential cross section for the trajectories of incident beam of electrons
  uniform per annulus $(\nu+1/2)d\nu$ may be obtained in a similar way
  as the classical one \eqref{eq:crossecCM} or \eqref{eq:crossecCM1}. 
By using \eqref{eq:scAngle_nu} and \eqref{eq:Pnddnu} and \eqref{eq:Qnddnu} we have
\begin{align}
 &\left ( d\theta_{\rm sc} \partial_{\theta_{\rm sc}}  + d\nu \partial_\nu
  \right ) \partial_{\nu} W_\theta(\theta_{\rm sc}, \nu) =
   d\nu \partial_\nu 2 \Im \psi(a),\notag\\
 &\frac{d\nu}{d\theta_{\rm sc}} = \frac{\partial_{\theta_{\rm sc}}
 \partial_{\nu} W_\theta(\theta_{\rm sc}, \nu) }{2 \Im \partial_a \psi(a)
  - \partial_{\nu}^2 W_\theta(\theta_{\rm sc}, \nu) },
    	\label{eq:dnudtetsc} \\
 &\sigma(\theta_{\rm sc}) =
  \frac{s}{\sin \theta_{\rm sc}}
     \left | \frac{d s}{d \theta_{\rm sc}} \right |
   = \frac{(\nu + \tfrac{1}{2})\hbar^2}{2 m E}
       \frac{1}{\sin \theta_{\rm sc}}
         \left | \frac{d \nu}{d \theta_{\rm sc}} \right |
   = \frac{1}{k^2} \frac{\nu + \tfrac{1}{2}}{\sin \theta_{\rm sc}}
         \left | \frac{d \nu}{d \theta_{\rm sc}} \right |.
       			\label{eq:dfcrosec_tr}
\end{align}
Parameter $\nu$ in the right hand side should be expressed
in terms of $\theta_{\rm sc}$ through  \eqref{eq:scAngle_nu}.

An example of the differential cross section vs the scattering angle is drawn in Fig.~\ref{fig:Difcrosect} in \S~\ref{sec:result}.

\subsection{ $dt/d\rho(\rho, E, \nu)$  for $-1/2 <\nu <0$  
  \label{Sec:dtdrnu<0} }
Numerical analysis of the following equations indicates the existence of the point 
 $\rho_0$ near the origin where $dt/d\rho(\rho_0, E, \nu)=0$ for  $-1/2 <\nu <0$.
 Therefore $\nu$ should be non-negative.
 It leads to the limiting scattering angle of a scattered electron with energy $E$
 by (\ref{eq:scAngle_nu}).
\begin{equation}
  \partial_{\nu} W_\theta(\theta_{\rm sc}, \nu)_{\nu=0} = - 2 \Im \psi(1+i \eta_s).
    \label{eq:sctangl-limit}
\end{equation}
 
\begin{align}
& 2E \frac{\partial}{\partial \rho} 
    \left (\frac{\partial W_r(r,E,\nu)}{\partial E} \right )
   =  2E \frac{\partial}{\partial \rho} \Im \left (\frac{\partial_E u}{u}
     \right )
  =-1 +  2E \frac{\partial}{\partial \rho} \Im \left (
     \frac{\partial_E \tilde{u}}{\tilde{u}}  \right )
    \notag\\
&= -1 + 4 E \Re \left (\frac{\partial_z\partial_E \tilde{u}}{\tilde{u}}
       - \frac{\partial_z \tilde{u}}{\tilde{u}}
               \frac{\partial_E \tilde{u}}{\tilde{u}} \right ). 
\end{align}
\begin{align}
&2E \partial_z\partial_E \tilde{u}  
 =  - e^{- i\pi b}
  \frac{\Gamma (1-b)\Gamma (b-a)}{ \Gamma (1-a)} \Biggl [ \left [
  i \eta_s \left \{ \psi(b-a) - \psi(1-a) \right \} + 1 + z - b \right ]
    \frac{a}{b} M(a+1, b+1, z)     \notag\\
 &+ a M(a, b, z) - \frac{i \eta_s}{b} \left (1 + a \frac{\partial}{\partial a} \right )
    M(a+1, b+1, z)_{z=i2\rho} \Biggr ]   \notag\\
  &{} - \Gamma(b) z^{-b} \Biggl \{ 
    1-b - i \eta_s \frac{\partial}{\partial A}
  + z \frac{\partial}{\partial z} \Biggr \} M(A,2-b,z)_{A=1+a-b, z=i2\rho} 
       \notag\\
   &{} + \Gamma(b-1) z^{1-b} \Biggl [    
      z \frac{\partial}{\partial z} M(A,2-b,z)_{A=1+a-b, z=i2\rho} 
    + A M(A,2-b,z)_{A=1+a-b, z=i2\rho}
    \notag\\
 & - \frac{i \eta_s}{2-b} \left (1 + A \frac{\partial}{\partial A} \right )
    M(A+1, 3-b, z)_{A=1+a-b, z=i2\rho} \Biggr ].
\end{align}

\section{ Temple orbit by Hamilton Jacobi equation  
   \label{Sec:CMTemple}}
As to the Coulomb scattering the Temple wave form is known in quantum
 mechanics\cite{MottMassey}.
In classical mechanics the corresponding orbit has not been shown 
to our knowledge. 
To compare the classical orbit and the wave trajectory described in the next 
section the Temple form solution of the Hamilton Jacobi equation of
 the Coulomb scattering will be investigated.

The Hamilton Jacobi equation is the same \eqref{eq:H-Jeq} but rewritten as
\begin{align}
&\frac{1}{2m}\left ( \nabla W_{\rm cl}  \right )^2 - \frac{Ze^2}{r} 
  = E.   \label{eq:H-JeqTmpl}
\end{align}
The Temple solution may be given by putting the characteristic function as
\[ W_{\rm cl}(x,y,z ) = W_{x, \rm cl}(x)  + W_{\zeta, \rm cl}(\zeta), \;
 \zeta=r-x.
\] 
Some calculation like 
\[ \partial_x \zeta = \frac{x}{r} - 1=- \frac{\zeta}{r},
  \partial_y \zeta = \frac{y}{r}, \partial_z \zeta = \frac{z}{r}
 \] will lead to
\begin{align*}
& \left ( \nabla W_{\rm cl}  \right )^2 
   = (\partial_x W_{\rm cl})^2 + (\partial_y W_{\rm cl})^2
   + (\partial_z W_{\rm cl})^2 \\
 &= \left (\frac{y^2+z^2}{r^2} + \frac{\zeta^2}{r^2} \right )
      (\partial_{\zeta} W_{\rm cl})^2
  - 2  \frac{\zeta}{r}  \partial_{\zeta} W_{\rm cl} \partial_x W_{\rm cl}
   + (\partial_x W_{\rm cl})^2
	\\
 &=2 \frac{\zeta}{r}(\partial_{\zeta} W_{\rm cl})^2
  - 2 \frac{\zeta}{r} \partial_{\zeta} W_{\rm cl} \partial_x W_{\rm cl}
   + (\partial_x W_{\rm cl})^2.
\end{align*}
 This suggests that the Hamilton-Jacobi equation~\eqref{eq:H-JeqTmpl}
 is separated
\begin{align}
&W_{x, \rm cl}(x) = px, \; p^2=2mE,  \label{eq:WclxT}  \\
& \left ( \frac{\mathrm{d} W_{\zeta, \rm cl}(\zeta)}{\mathrm{d} \zeta}
    \right )^2
    -  p \frac{\mathrm{d} W_{\zeta, \rm cl}(\zeta)}{\mathrm{d} \zeta}
    - \frac{mZe^2}{\zeta} = 0.  
\end{align}
Some more calculation and integration gives rise to
\begin{align}
&W_{\rm cl, \zeta}(\zeta)^\pm
  = \frac{p}{2}\zeta \pm \frac{p}{2} \sqrt{\zeta \left (
     \zeta + \frac{2Ze^2}{E} \right )} \mp \frac{pZe^2}{2E} 
     \log \left |\frac{\sqrt{\zeta + \tfrac{2Ze^2}{E} } - \sqrt{\zeta}}{
       \sqrt{\zeta + \tfrac{2Ze^2}{E} } + \sqrt{\zeta}} \right |.
       \label{eq:WclzetaT} 
\end{align}
\eqref{eq:WclxT} and \eqref{eq:WclzetaT} leads to for $\zeta=r - x \to \infty$
\begin{align}
 \frac{\partial W_{\rm cl, \zeta}(\zeta)^\pm }{\partial E}
   +\frac{\partial W_{x, \rm cl}(x) }{\partial E} \to 
 \sqrt{\frac{m}{2E}} \left ( \genfrac{.}{.}{0pt}{}{r}{x}
  \mp \frac{Ze^2}{2E} \log \frac{2E\zeta}{Ze^2} \right )
  \label{eq:TimeCMT0}
\end{align}
 This corresponds to the time elapse of the particle in the Coulomb field \eqref{eq:TimeCL}.
Equations~\eqref{eq:WclxT} and \eqref{eq:WclzetaT} could not, however, lead to the 
orbit. \\
To derive the orbit and dynamics in one way or another  let us rotate the coordinates 
$(x,y)$ to $(x',y')$ with an arbitrary angle $\varphi$

\begin{equation}
 \binom{x'}{y'} 
	 = \begin{pmatrix}
	\cos \varphi & - \sin \varphi \\
	\sin \varphi &  \cos \varphi
    \end{pmatrix} \binom{x}{y}.
    \label{eq:rotate}
\end{equation}
Since $(\partial_x W_{\rm cl})^2+(\partial_y W_{\rm cl})^2 
=(\partial_{x'} W_{\rm cl})^2+(\partial_{y'} W_{\rm cl})^2$ and
$r'\equiv \sqrt{x'^2+y'^2+z^2}=r$,  Hamilton Jacobi equation is written as
\[ \frac{1}{2m} \left \{ (\partial_{x'} W_{\rm cl})^2
  +(\partial_{y'} W_{\rm cl})^2
  +(\partial_{z} W_{\rm cl})^2 \right \} - \frac{Ze^2}{r} = E.
\]
Therefore we have the characteristic function dependent on
 $\zeta'=r - x', x'$, with $p=\sqrt{2mE}$,
\begin{align}
&W_{\rm cl}(x,y,z;\varphi, E)=W_{\rm cl,x'}(x' ;  E)
    + W_{\rm cl,\zeta'}(\zeta' ;  E),   \label{eq:WclTmp}\\
& W_{\rm cl,x'}(x' ;  E) =\sqrt{2mE} x',   \label{eq:dWclTmpxp} \\
&W_{\rm cl,\zeta'}(\zeta' ;  E)
 =\{W_{\rm cl,\zeta'}^+(\zeta';  E), W_{\rm cl,\zeta'}^-(\zeta';  E)\},
   \notag\\
 &W_{\rm cl,\zeta'}^\pm(\zeta' )
 =W_{\rm cl,\zeta'}^\pm(r - (\cos \varphi x - \sin \varphi y); E)\notag\\
 &= \sqrt{\frac{mE}{2}}\zeta' \pm \sqrt{\frac{mE}{2}} \sqrt{\zeta' \left (
     \zeta' + \frac{2Ze^2}{E} \right )} \mp Ze^2\sqrt{\frac{m}{2E}} 
    \log \left |\frac{\sqrt{\zeta' + \tfrac{2Ze^2}{E} } - \sqrt{\zeta'}}{
     \sqrt{\zeta' + \tfrac{2Ze^2}{E} } + \sqrt{\zeta'}} \right |.
     \label{eq:WclTmpzeta}
 \end{align}

The orbit and the dynamics should be given by
\begin{align}
 & \partial_{\varphi} W_{\rm cl}(x,y,z;\varphi, E)
  = \partial_{\varphi} \left ( W_{\zeta'}(\zeta'; E)^\pm + W_{x'}(x'; E)
  \right )
   = y' \left ( \frac{\rm d}{\rm d \zeta'} W_{\zeta'}(\zeta'; E)^\pm
     - \frac{\rm d}{\rm d x'}W_{x'}(x'; E )   \right ) \notag\\
 &\qquad=  y' \left ( - \sqrt{\frac{mE}{2}}
   \pm \sqrt{ \frac{mE}{2} + \frac{mZe^2}{\zeta'}} \right )
 = y_0 (\text{constant}), 	\\
 & \partial_{E} W_{\rm cl}(x,y,z;\varphi, E)
  = \partial_{E} \left ( W_{\zeta'}(\zeta'; E )^\pm + W_{x'}(x'; E ) \right )
  = t + t_0 (\text{constant}).
   \label{eq:CMTmpdyn}
\end{align}
Here use has been made of
\[ \partial_{\varphi} \zeta' = - \partial_{\varphi} x' 
    = \sin \varphi x + \cos \varphi y = y'.
  \]  

Let $x=r \cos \theta, y=r \sin \theta$. 
For the scattering state that the incident electron from $x =\infty,
 py=ps =L$(constant) is scattered by the Coulomb potential the incident
 characteristic function is $W_{\zeta'}^{-}(\zeta') + W_{x'}(x' )$ and
 the scattered one is  
$W_{\zeta'}^{+}(\zeta') + W_{x'}(x' )$ with $\varphi \to 0$.
\begin{align}
 & \partial_{\varphi} \left ( W_{\zeta'}(\zeta' )^{-} + W_{x'}(x' )
             \right )_{\varphi=0}
     = y \left ( \partial_{\zeta'} W_{\zeta'}(\zeta' )^{-}
                 - \partial_{x'}W_{x'}(x' ) \right ) 
             \notag\\
 &\qquad= y \left (- \sqrt{ \frac{mE}{2}}
  - \sqrt{ \frac{mE}{2} + \frac{mZe^2}{\zeta}}  \right ) = - L,
       \label{eq:CMTmpinc} \\
 & \partial_{\varphi} \left ( W_{\zeta'}(\zeta' )^+ + W_{x'}(x' )
             \right )_{\varphi=0}      
  = y \left ( - \sqrt{ \frac{mE}{2}}
    + \sqrt{ \frac{mE}{2} + \frac{mZe^2}{\zeta}}  \right )
  = - L .    \label{eq:CMTmpsct}
\end{align}
For $r \to \infty, \theta \to \pi, \zeta=r - x \to \infty$ 
\[ \partial_{\varphi} \left ( W_{\zeta'}(\zeta' )^- + W_{x'}(x' )
             \right )_{\varphi=0} \to - \sqrt{ 2mE} y = - p s = - L. 
\]
Here, $p$ is the momentum, $L$ is the angular momentum,
 and $s$ is the impact parameter.
 From 
\[ \partial_{\varphi} \left ( W_{\zeta'}(\zeta' )^+ + W_{x'}(x' )
             \right )_{\varphi=0} = - ps
\]
for $r \to \infty, \theta \to \theta_\text{sc}$ cf.\eqref{eq:etasrho}
\[ \sqrt{ \frac{mE}{2}} \frac{Ze^2}{E} \frac{y}{\zeta} 
 =\frac{p}{2} \frac{Ze^2}{E} \frac{\sin \theta_\text{sc}}{1 - \cos \theta_\text{sc}}
 = - ps, \text{ or } \tan \frac{\theta_\text{sc}}{2} = - \frac{Ze^2}{2Es}
  = - \frac{Zme^2}{pL} = - \frac{\eta_s}{\ell}.
\]
 
 In what follows $p = k \hbar$ and $L=\ell \hbar $ are used.
From \eqref{eq:CMTmpinc} for $px/\hbar = k x \to -\infty$, $py = ps=\ell \hbar$
 is attained
 and the orbit equation is explicitly written from \eqref{eq:CMTmpinc} and 
 \eqref{eq:CMTmpsct} as
\begin{align}
 \frac{s}{r} =\frac{Ze^2}{2Es} (1 + \cos \theta) +  \sin \theta,
  \text{ or, }
 \frac{\ell}{kr } =\frac{\eta_s}{\ell} (1 + \cos \theta) +  \sin \theta.
     \label{eq:CMTmpObt}
\end{align}
This is a typical hyperbolic curve of orbit in the Coulomb potential.

That the orbit obtained from \eqref{eq:CMTmpsct} should accord with this 
equation leads to $y_0 = -L= - ks \hbar$. 
By taking $kr \to \infty $ in  \eqref{eq:CMTmpsct} the scattering 
angle $\theta_{\rm sc}$ is obtained:
\begin{align}
 \theta_{\rm sc}= 2 \arctan \left ( \frac{Zme^2}{- pL} \right )
   = - 2 \arctan \left ( \frac{\eta_s}{ks} \right ).
     \label{eq:CMTmpsctangl}
\end{align}
This is equivalent to \eqref{eq:sctangle_cls}.  
The returning point where the incident orbit \eqref{eq:CMTmpinc} transfer
 to the scattering orbit  \eqref{eq:CMTmpsct} is $y=0, \theta_\text{ret}=0$ for $Z>0$.
 For $Z<0$ it is given by the conditions $E \zeta + 2 Ze^2 =0$ and $py = 2ps$, 
 and thus
 $\theta_\text{ret} = 2 \arctan \left ( -Z e^2/(E s) \right ), r_\text{ret} = 2s/\sin \theta_\text{ret}  $.

The time elapse of the motion for the incident orbit \eqref{eq:CMTmpinc} 
and the scattering orbit  \eqref{eq:CMTmpsct}  are
\begin{align}
 & ( t + t_0 )^\text{in} = \partial_E \left ( W_{\zeta'}(\zeta' )^- + W_{x'}(x' )
             \right )_{\varphi=0}    \notag\\
  &= \sqrt{\frac{m}{8E}} \left \{ \zeta - \left (
    \sqrt{ \zeta \left ( \zeta +\frac{2Ze^2}{E} \right ) }
   - \frac{Ze^2}{E} \log \left |
     \frac{\sqrt{\zeta} +\sqrt{\zeta +\frac{2Ze^2}{E}} }{
    \sqrt{ \zeta} -\sqrt{ \zeta +\frac{2Ze^2}{E}} } \right | \right )
   \right \}  +  \sqrt{\frac{m}{2E}} x,  
      \label{eq:CMTmptimeinc}     \\
&  ( t + t_0 )^\text{sc} = \partial_E \left ( W_{\zeta'}(\zeta' )^+ + W_{x'}(x' )
             \right )_{\varphi=0}    \notag\\
 &= \sqrt{\frac{m}{8E}} \left \{ \zeta +  \left (
    \sqrt{ \zeta \left ( \zeta +\frac{2Ze^2}{E} \right ) }
   - \frac{Ze^2}{E} \log \left |
     \frac{\sqrt{\zeta} +\sqrt{\zeta +\frac{2Ze^2}{E}} }{
    \sqrt{ \zeta} -\sqrt{ \zeta +\frac{2Ze^2}{E}} } \right | \right )
   \right \}   
  +  \sqrt{\frac{m}{2E}} x.     \label{eq:CMTmptimesct}  
\end{align}
From \eqref{eq:CMTmptimeinc} for $kr \to \infty, \theta \to \pi$,
\begin{align}
( t + t_0 )^\text{in} \simeq  \sqrt{\frac{m}{2E}} x 
  + \sqrt{\frac{m}{8E}} \frac{Ze^2}{E} \log \frac{2E\zeta}{Ze^2}
  \simeq \frac{mx}{p}
  + \sqrt{\frac{m}{8E}} \frac{Ze^2}{E} \log \frac{4E|x|}{Ze^2}.
 \label{eq:t-TmplCM-incApp}
\end{align}
From \eqref{eq:CMTmptimesct} for $kr \to \infty, \theta \to \theta_\text{sc}$,
\begin{align}
( t + t_0 )^\text{sc} \simeq \sqrt{\frac{m}{2E}} r
   - \sqrt{\frac{m}{8E}} \frac{Ze^2}{E} \log \frac{2E r(1 - \cos \theta_\text{sc})}{Ze^2}.
  \label{eq:t-TmplCM-sctApp}
\end{align}
These are comparable to equations \eqref{eq:timeelapse} for the m-trajectory and
 the following eqs. \eqref{eq:t-Tmpl-incApp} and \eqref{eq:t-Tmpl-sctApp} for the Temple
 m-trajectory.\\
The dynamics of \eqref{eq:CMTmpinc}, \eqref{eq:CMTmpsct}, \eqref{eq:CMTmptimeinc} and \eqref{eq:CMTmptimesct} obtained by Hamilton-Jacobi equation in Temple coordinates are equivalent to the dynamics of \eqref{eq:CMobtinc}, \eqref{eq:CMobtsct} and \eqref{eq:TimeCL} obtained by Hamilton-Jacobi equation in the spherical polar coordinates, both  in the Cartesian coordinates system. 

Examples of the orbit and time elapse of a scattering electron for $E=20$eV are shown in Fig.~\ref{fig:4}.  and Fig.~\ref{fig:5}.

\section{ Temple mode trajectory from the wave function  \ 
  \label{Sec:QMTemple}}
A solution of the Schr\"{o}dinger equation has been given
 by Temple \cite{Temple} and rewritten in Mott \& Massey~\cite{MottMassey}
\begin{align}
 &\nabla^2 \psi(x,y,z) + \frac{2m}{\hbar^2} \left (
 E + \frac{Z e^2}{r} \right ) \psi(x,y,z)
 = k^2 \left ( \frac{1}{k^2} \nabla^2 + 1
   +  \frac{2 \eta_s}{k r} \right )\psi(x,y,z) = 0.
    \label{eq:Schroedeq}
		\\
 & k \equiv \frac{\sqrt{2mE}}{\hbar}, \;  \eta_s = \frac{ Z me^2}{\hbar^2 k}.
       \; \text{ cf. }\eqref{eq:etasrho}     \notag 
\end{align}
Temple form solution is obtained by setting
\begin{align}
&\psi(x,y,z ) = e^{ikx} F(\zeta ) , \quad \zeta=r - x, \notag\\
& \left (ik\zeta \frac{\text{d}^2}{\text{d} (ik\zeta)^2} + 
       (1 - ik\zeta) \frac{\text{d}}{\text{d}ik\zeta} - i\eta_s 
         \right ) F(\zeta) =0.
    \label{eq:Fzetaeq}
\end{align}
The linearly independent solutions are the Kummer functions~\cite{AbraSteg},
 \begin{align}
 &F(\zeta) = \left \{ M(i\eta_s, 1, ik\zeta), U(i\eta_s, 1, ik\zeta) \right \}. 
     \label{eq:Fialfa}  \\
 &M(a,1,z)=\sum_{n=0}^\infty \frac{(a)_n z^n}{n! n!},\;
  U(a,1,z)= - \frac{1}{\Gamma(a)} \left [ M(a,1,z) \log z + \psi(a)
   - 2\psi(1) \right ], 
\end{align}
 where $a=i\eta_s$ or $1- i\eta_s$ and $z=ik\zeta$ or $- ik\zeta $.

For $|z|$ large, $a,b$ fixed  asymptotic expressions for $ U(a,b,z)$,~\cite{AbraSteg},
\begin{align}
 &U(a,b,z) \sim  z^{-a}
  \left (1 - \frac{a(1+a-b)}{z} \right ) + z^{-a} O(|z|^{-2}).
  \quad{} -\frac{3\pi}{2} < \arg(z) < \frac{3\pi}{2}
              \label{eq:asymptU}
\end{align}
The wave functions having the similar phase to characteristic functions of Temple 
form of classical mechanics should be sought.
Define the incident and scattering functions from  
  \eqref{eq:Fzetaeq} and \eqref{eq:Fialfa}
\begin{align}
  &\psi_{\rm in}(x,y,z) \equiv e^{-\pi\eta_s/2} e^{ikx} U(i\eta_s,1, ik\zeta)
 \simeq e^{i(kx - \eta_s \log(k\zeta)}, 
   \label{eq:psi_in} \\
 &\psi_{\rm sc}(x,y,z) \equiv e^{\pi\eta_s/2}\Gamma(1-i\eta_s)
   e^{ikx} M(i\eta_s,1, ik\zeta) - \psi_{\rm in}(x,y,z)   \notag\\
 &\quad = - \frac{e^{-\pi \eta_s/2}\Gamma(1 - i\eta_s)}{\Gamma(i \eta_s)}
    e^{ ikr} U(1 - i\eta_s,1,- ik\zeta) 
  \simeq \frac{-i \Gamma(1-i\eta_s)}{\Gamma(i\eta_s)}
  \frac{e^{i kr + i\eta_s \log(k\zeta) }}{ k\zeta}.
     \label{eq:psi_sc}
\end{align}

 Each wave function in the remote region $kr \to \infty$ is shown in the
 following of the signature $\simeq$.
Thus 
$\psi_{\rm in}(x,y,z)$ shows the unit incident plane-like wave in the Coulomb field.
$\psi_{\rm sc}(x,y,z)$ represents the diverging scattered wave.
 These functions have a singular point  $kr=0$.
But the sum function $\psi_{\rm in}(x,y,z) + \psi_{\rm sc}(x,y,z)$ is
 regular everywhere.

 These functions cannot give rise to the trajectory like classical 
mechanics of the preceding section.
To derive the trajectory let rotate coordinate $(x, y)$ to  $(x', y')$
 according to \eqref{eq:rotate}.
We consider the wave in the coordinates $(x', y')$ rotated by an arbitrary
 angle $\varphi$ from the coordinates $(x,y)$.
The Schr\"{o}dinger equation is rewritten as
\begin{align}
 &\nabla^{\prime 2} \psi(x',y',z) + \frac{2m}{\hbar^2} \left (
 E + \frac{Z e^2}{r} \right ) \psi(x',y',z)
 = k^2 \left ( \frac{1}{k^2} \nabla^{\prime 2} + 1
   +  \frac{2 \eta_s}{k r} \right )\psi(x',y',z) = 0. 
    \label{eq:Schroedeqprm} \\
 &\nabla^{\prime 2}=\partial_{x'}^2 + \partial_{y'}^2 + \partial_{z}^2
    = \partial_{x}^2 + \partial_{y}^2 + \partial_{z}^2=\nabla^{ 2}. \notag
\end{align}
We get the expected functions
\begin{align}
  &\psi_{\rm in}(x',y',z) = e^{-\pi\eta_s/2} e^{ikx'} U(i\eta_s,1, ik\zeta'),
   \label{eq:psi_inprm} \\
 &\psi_{\rm sc}(x',y',z) =
  - \frac{e^{-\pi \eta_s/2}\Gamma(1 - i\eta_s)}{\Gamma(i \eta_s)}
    e^{ ikr} U(1 - i\eta_s,1,- ik\zeta'). 
     \label{eq:psi_scprm}
\end{align}
Here, $\zeta'=r - x'$.
By using the relations
\begin{align}
 &\frac{\partial}{\partial \varphi} \psi_{\rm in}(x',y',z)
  =- ik y' \{ \psi_{\rm in}(x',y',z) +
  i\eta_s e^{-\pi\eta_s/2} e^{ik x'}  U(i\eta_s+1, 2, i k \zeta') \}, \\
 &\frac{\partial}{\partial \varphi} \psi_{\rm sc}(x',y',z)
 = - \frac{e^{-\pi \eta_s/2}\Gamma(2 - i\eta_s)i k y'}{\Gamma(i \eta_s) }
     e^{ ikr} U(2 - i\eta_s,2,- ik\zeta'),  
\end{align}
the mode trajectory equations (named as Temple m-trajectory) are given by
\begin{align}
 &C_1= \Im \left ( \frac{\partial_\varphi \psi_{\rm in}(x',y',z)}{
   \psi_{\rm in}(x',y',z) } \right )_{\varphi=0}
  = - k y + ky\eta_s  \Im  \left ( \frac{ U(i\eta_s+1, 2, i k \zeta)}{ 
      U(i\eta_s, 1, i k\zeta)} \right ) \simeq - ky,  
                \label{eq:trjTmp_in}\\
 &C_2=\Im \left ( \frac{\partial_\varphi \psi_{\rm sc}(x',y',z)}{
   \psi_{\rm sc}(x',y',z) } \right )_{\varphi=0}
 =   k y \Re \left ( (1 - i\eta_s) 
     \frac{ U(2 - i\eta_s,2,- ik\zeta)}{U(1 - i\eta_s,1,- ik\zeta)} \right )  
   \simeq \eta_s \cot \frac{\theta}{2}. \label{eq:trjTmp_sc}
\end{align}
Here use has been made of \eqref{eq:asymptU}.
The function following $\simeq$ shows the approximate one for $kr \to \infty$.
Corresponding characteristic functions indicate that $C_1=- ks =- \ell$ and
 $C_2=- ks=- \ell$ where $s=y$ at $r \to \infty$ is the impact parameter.
 The scattering angle is given by
 $\theta_{\rm sc}=-2 \arctan (\eta_s/ks)=-2 \arctan (\eta_s/\ell)$
 which is equal to that of classical mechanics \eqref{eq:CMTmpsctangl}.

The time dependence of the mode trajectory is given by the relations
\begin{align}
& k \Im \left (
    \frac{\partial_k \psi_{\rm in}(x',y',z)}{\psi_{\rm in}(x',y',z)}
    \right )_{\varphi=0}
   =   kx + \Im \left \{
   - i \eta_s \frac{\partial_a U(a,1,ik\zeta)}{ U(a,1,ik\zeta)}
   \Bigg|_{a=i\eta_s}
  + \eta_s k \zeta \frac{ U(i\eta_s+1,2,ik\zeta)}{U(i\eta_s,1,ik\zeta)}
  \right \},   \notag\\
 & (t + t_0)_{\rm in} =  \hbar \Im \left (
    \frac{\partial_E \psi_{\rm in}(x',y',z)}{\psi_{\rm in}(x,y,z)} \right )_{\varphi=0}
   = \frac{\hbar k}{2E} \Im \left (
    \frac{\partial_k \psi_{\rm in}(x',y',z)}{\psi_{\rm in}(x,y,z)} \right )_{\varphi=0},
  \label{eq:tTmp_inc} \\
 &\qquad \text{where $x,y$ are correlated by \eqref{eq:trjTmp_in}, }
   \notag\\
& k \Im \left (
    \frac{\partial_k \psi_{\rm sc}(x',y',z)}{\psi_{\rm sc}(x',y',z)}
     \right )_{\varphi=0}
   =  kr + \eta_s \Re \left \{ \psi(1-i\eta_s) + \psi(i\eta_s) 
   + \frac{\partial_a U(a,1,-ik\zeta)}{ U(a,1,-ik\zeta)}
   \Bigg|_{a=1-i\eta_s} \right \}  \notag\\
 &\qquad{}+ k\zeta \Re \left \{
  \frac{(1-i\eta_s) U(2-i\eta_s,2,-ik\zeta)}{ U(1-i\eta_s,1,-ik\zeta)}
    \right \}, \notag\\
& (t + t_0)_{\rm sc} = \frac{\hbar k}{2E} \Im \left (
    \frac{\partial_k \psi_{\rm sc}(x',y',z)}{\psi_{\rm sc}(x,y,z)} \right )_{\varphi=0},
  \label{eq:tTmp_sct}  \\
& \qquad \text{where $x,y$ are correlated by \eqref{eq:trjTmp_sc}. }
   \notag
\end{align}
 For $r \to \infty, \theta \to \pi, x \to -\infty$
\begin{align}
 (t + t_0)_{\rm in} \simeq  \frac{\hbar }{2E} \Im \left \{ ikx
   + i \eta_s \log (ik\zeta)+ \eta_s k \zeta /(ik\zeta) \right \}
  = \frac{\hbar }{2E} \left \{ kx
   +  \eta_s \log (k\zeta) - \eta_s \right \}.  \label{eq:t-Tmpl-incApp}
\end{align}
For $r \to \infty, \theta \to \theta_\text{sc}$
\begin{align}
 &(t + t_0)_{\rm sc} \simeq  \frac{\hbar }{2E} \Re \left [ kr
  + \eta_s \left \{ \psi(1-i\eta_s) + \psi(i\eta_s) - \log(-ik\zeta)
  + k\zeta \frac{1-i\eta_s}{-ik\zeta} \right \} \right ]  \notag\\
 &= \frac{\hbar }{2E}  \left [ kr  - \eta_s \log(k\zeta) 
  + \eta_s \Re \left \{ \psi(1-i\eta_s) + \psi(i\eta_s)  \right \} 
  + \eta_s^2 \right ].  \label{eq:t-Tmpl-sctApp}
\end{align}
These asymptotic times for the remote point from the origin of the potential
 correspond well to the values in the classical mechanics
 \eqref{eq:t-TmplCM-incApp} and \eqref{eq:t-TmplCM-sctApp}.

Examples of Temple mode trajectory and dynamics are shown in Fig.~\ref{fig:4} and Fig.~\ref{fig:5}.

\subsection{Cross section}
In the Temple coordinate system $x, \zeta, z$ the impact parameter $s$, or 
 angular momentum $ps = \hbar ks = \hbar \ell$ and the energy
 $E = (\hbar k)^2/(2m)$ determine the dynamics of the scattering electron.
The scattering angle $\theta_\text{sc} = -2 \arctan(Zme^2/(2Es))$ is correlated
 in the same way in both classical mechanics \eqref{eq:CMTmpsctangl} and
 wave mechanics \eqref{eq:trjTmp_sc}.
Therefore the dependence of the cross section on the scattering angle is 
 the same as \eqref{eq:crossecCM1}  
\begin{align}
   \sigma( \theta_\text{sc})= \frac{s}{\sin \theta_\text{sc}} \left |
    \frac{\text{d}s}{\text{d}\theta_\text{sc}} \right |
  = \frac{\eta_s^2}{4k^2} \csc^4 \left (\frac{\theta_\text{sc}}{2} \right ).
  \label{eq:crossect_templ}
\end{align}

\section{ Numerical results and discussions}\label{sec:result}
 In the remote region from the origin the m-trajectory in the spherical polar or Temple coordinates is very similar to the corresponding classical orbit but in the region near the neighbourhood of the center of the potential the trajectory function is too complex to see the characteristics of the motion.
It is necessary to analyse numerically the trajectory near the origin in detail to judge
 the validity of the m-trajectory.

\begin{figure}[htbp]
\begin{center}
  \includegraphics[width=7.8cm]{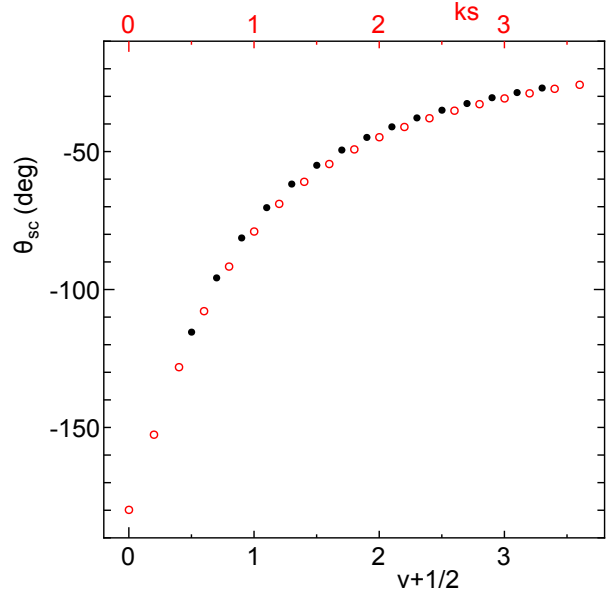} 
\end{center}
\caption{ Scattering angle of an incident electron with $E=20$eV, $Z=1$
  as a function of $ks$  for the classical orbit or Temple
 trajectory (red circle) and as that of $\nu+1/2$ 
 for the m-trajectory in the spherical polar coordinates (black dot).
\label{fig:2} }
\end{figure}

 Scattering angle $\theta_\text{sc}$ as a function of the impact parameter $s$
  of an incident electron with energy $20$eV for $Z=1$
 for the classical orbit (red circle)~\eqref{eq:CMTmpsctangl},
 or Temple m-trajectory~\eqref{eq:trjTmp_sc}  and
 for the m-trajectory in the spherical polar coordinates (black dot) \eqref{eq:scAngle_nu}
   is shown in Fig.~\ref{fig:2}. 
Figure~\ref{fig:2} indicates that $\nu+1/2$ in the m-trajectory in the spherical polar coordinates has the same role as $ks=\ell$ in classical mechanics. Thus the differential cross section for the m-trajectory should be given by \eqref{eq:dfcrosec_tr}.

\begin{figure}[htbp]
\begin{center}
  \includegraphics[width=7.8cm]{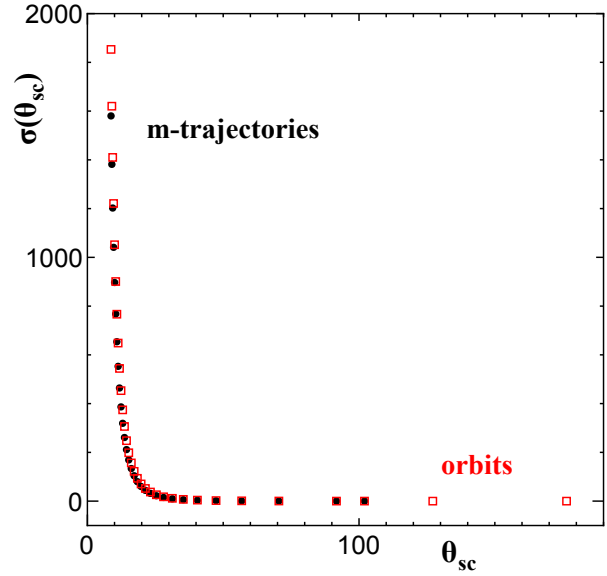} 
\end{center}
\caption{ Differential cross section as a function of the scattering
 angle of an incident beam with $E$=2000eV, Z=6 for the classical orbits or
 Temple trajectories (red box) and for the m-trajectories (black dot).}
\label{fig:Difcrosect}
\end{figure}

An example of the differential cross section of an electron beam of 
 incident energy 2000eV by nuclear charge $Z=6$
  for the classical orbit and that for the m-trajectory are shown
  in Fig.~\ref{fig:Difcrosect}.  
The figures indicate that the similarity between the cross sections
 of the classical orbits and the m-trajectories  seems complete for any
 values of charge $Z$ and energy $E$. 
 The difference between them is the existence of the limit of the scattering angle
 given by \eqref{eq:sctangl-limit} in the mode trajectory
 while $\theta_\text{sc} \to -\pi$ for $ks \to 0$ in the classical orbit. 

\begin{figure}[htbp]
\begin{center}
 \includegraphics[width=6.9cm,clip]{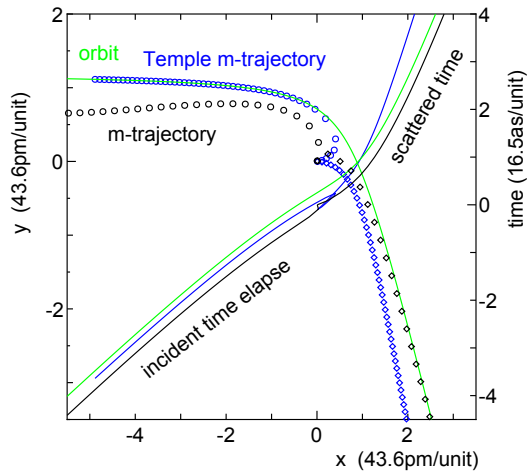} 
\end{center}
 \caption{\small Dynamics of a scattering electron near the center of the hydrogen
  with Energy=20eV, $\nu=0.629 $, m-trajectory  (black); 
 Temple m-trajectory (\textcolor{blue}{blue}) and
 the classical orbit (\textcolor{green}{green line})  with $ks=1.2, s=52.4\text{pm}$.  
Parameters $\nu$ and $ks$ have been set so that the scattering angle is all the same $\theta_\text{sc}=-69^\circ$. 
\label{fig:4} }
\end{figure}

\begin{figure}[htbp]
\begin{center}
 \includegraphics[width=6.9cm,clip]{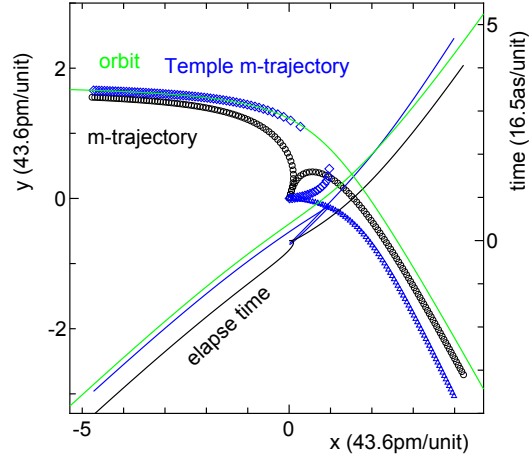} 
\end{center}
 \caption{\small Another example of dynamics of a scattering electron with $E=20$eV,
 $\nu  = 1.2, ks=1.79$, $\theta_\text{sc}=-49.5^\circ$. Other specifications are the same
 as Fig.~\ref{fig:4}. 
   \label{fig:5} }
\end{figure}

Examples of the m-trajectory in the spherical polar and Temple coordinates
 for some parameters of $\nu, L=\hbar k s=\hbar \ell,$
 $E=20$eV, $Z=1$ and scattering angle $\theta_\text{sc}$
 with the corresponding orbit of classical mechanics are shown in
  Figs.~\ref{fig:4} and~\ref{fig:5}.
In the remote region from the origin of the potential the m-trajectory and
 classical orbit correspond well and thus the behavior in the neighbourhood
 of the origin are to be investigated in detail.

Figures~\ref{fig:4} and~\ref{fig:5} by a detailed numerical calculation
 indicate a lack of the Temple m-trajectory and an irregular time elapse in a small region near the origin, although in the region outside of the small region the trajectory is almost equal to the corresponding orbit of classical mechanics.

The m-trajectory in the polar coordinate system is sound.  But for the m-trajectory
 with $-1/2 < \nu <0$ which might be considered to exist 
the time elapse in the neighbourhood of the origin 
 has a peculiar character. It has been clarified by numerical calculation. 
It suggests that $\nu$ should be non-negative.

\newpage
\section{Conclusion}\label{Sec:Conclusion}
The Schr\"{o}dinger wave equation can describe the mode trajectory and dynamics 
of an electron of the Coulomb scattering in the space-time if the wave
 function is well manipulated to treat the motion of the particle.
It is similar to the classical motion especially in the remote region from
 the origin of the potential as it should be.

The mode trajectory in Temple coordinates is not complete specifically in the near region around the origin of the potential.

The mode trajectory and dynamics in the spherical polar coordinates for $\nu \geq 0$ and $E >0$  is sound. 
It may be proved  by an experiment to show the existence of the limiting scattering angle \eqref{eq:sctangl-limit}.

\end{document}